 \documentclass{article}
\usepackage[labelfont=bf]{caption}
\usepackage{cite}
\usepackage{adjustbox}
\usepackage{hyperref}
\usepackage{tabularx}
\usepackage{placeins}
\usepackage[a4paper,top=1cm,bottom=2cm,left=3cm,right=3cm,marginparwidth=1.75cm]{geometry}
\usepackage{parskip}
\usepackage{etoolbox}
\usepackage{graphicx}
\usepackage{amsmath}
\usepackage{caption} 
\usepackage{lipsum}  
\usepackage{float}
\usepackage{authblk}
\hypersetup{
    colorlinks=true,
    linkcolor=black,
    filecolor=black,      
    urlcolor=black,
    citecolor=black
    }
\patchcmd{\thebibliography}{\chapter*}{\section*}{}{}
\title{\vspace{-1.5cm} Through the bottle authentication of red wine using near-IR fluorescence spectroscopy }
\author[1,2]{An\'e Kritzinger}
\author[1]{Ralf Mouthaan}
\author[2]{Graham D. Bruce}
\author[3]{Eric Wilkes}
\author[1,2]{Kishan Dholakia}
\affil[1]{Centre of Light for Life and School of Biological Sciences, Adelaide University, Adelaide, 5005, Australia}
\affil[2]{SUPA, School of Physics and Astronomy, University of St Andrews, North Haugh, St Andrews, Fife, KY16 9SS, UK}
\affil[3]{The Australian Wine Research Institute, Hartley Grove, Urrbrae, Adelaide, 5064, Australia}

\date{}

\begin{document}

% \twocolumn[
%   \begin{@twocolumnfalse}
\maketitle

\begin{abstract}

A major unaddressed challenge for food science remains the accurate characterisation of contents in sealed containers with a non-invasive method. This issue is particularly pressing for tackling fraud in the red wine industry, valued at billions of dollars globally, where product authenticity, brand reputation, and consumer trust are paramount. Whilst many techniques exist for authenticating wine externally, to date performing accurate classification of the contents within unopened bottles remains elusive. Using only a single near-infrared optical excitation source operating at a wavelength of 785~nm, in combination with a bespoke geometry to circumvent the confounding signal of the glass, we demonstrate that through-bottle fluorescence spectra can distinguish between twenty different red wines in their original, intact bottles. All twenty wine bottles were correctly classified with linear discriminant analysis (LDA) and principal component analysis (PCA) revealed strong varietal grouping. This non-invasive and rapid technique has the potential to enable on-site, routine wine authentication to combat the growing issue of wine fraud. The geometry itself is applicable across multiple fields for the analysis of other high-value products through their packaging, where authenticity verification is critical.

\end{abstract}
% \textbf{Keywords:} Fluorescence spectroscopy, red wine, through-the-bottle, wine authentication, non-invasive.\\

%   \end{@twocolumnfalse}
% ]

\section{Introduction}

% One of the major challenges in food science is the acquisition of information directly through containers without opening the contents. Such a capability is a burgeoning requirement for authentication provenance and reducing fraud. Such an approach would supersede approaches with digital labelling or otherwise and be of immense value in areas such as the wine and liquor industry, studies of olive oil and other commodities such as baby milk powder.  

% RPM: High-value food items are particularly susceptible to food fraud such as mislabelling, substitution, dilution or adulteration. These malpractices have impacted on almost every corner of the food sector, including dairy, meat, fish, spice, alcoholic and non-alcoholic beverages. The global wine market, valued at USD 336 billion in 2023 (Euromonitor), is particularly vulnerable to fraudulent activities. The monetary value and reputation of a bottle of wine depends on several factors, including the varietal, provenance, vintage, and brand of the wine. Wine fraud includes counterfeiting labels, adulteration with prohibited chemicals, and substituting grapes with cheaper varietals or with those from different geographical origins or vintages. A 2017 study estimated that Australia lost several hundred million dollars that year due to fraud \cite{mcleod2017counting}. More recently, in 2024, an international fraud ring operating in France and Italy was dismantled for selling cheap wine as famous French vintages, with prices reaching up to €15,000 (A\$24,000) per bottle (\url{https://www.bbc.com/news/articles/cvg3jzzjg3po}). 

A major challenge in food science is the acquisition of information through unopened containers. This capability is a burgeoning requirement for authenticating provenance and reducing fraud. An approach that analyses the contents directly through dark or tinted glass containers would supersede external anti-fraud measures that can be prone to tampering. Non-invasive approaches offer immense value in areas such as the wine and beverage industry, studies of olive oil or other commodities such as pharmaceutical products or perfumes.

The global wine market, valued at US\$336 billion in 2023 (Euromonitor), is particularly vulnerable to fraudulent activities. The monetary value and reputation of a bottle of wine depend on several factors, including the varietal, provenance, vintage, and brand of the wine. Wine fraud includes counterfeiting labels, adulteration with prohibited chemicals, and substituting grapes with cheaper varietals or with those from different geographical origins or vintages. As with all illegal markets, knowing the exact size or financial impact of wine fraud is difficult. For the past decade, however, the phrase ``20\% of all wine bottles are fake" has frequently been asserted by wine experts \cite{lecat2017fraud}. 

Even the limited available data on wine fraud paints a troubling picture worldwide \cite{lin2021fraud}. For instance, a 2017 study estimated that Australia alone lost A\$303 million to wine fraud that year \cite{mcleod2017counting}. In Italy, annual losses are reported to be as high as €406 million \cite{romano2021sam}. In 2010, a prominent counterfeit lawyer revealed that over 70\% of the wine bottles he examined in China were fake \cite{bull2016grape}. More recently, in 2024, an international fraud ring operating in France and Italy was dismantled for selling cheap wine as famous French vintages, with prices reaching up to €15,000 per bottle \cite{bbc_wine}, wile in the UK in 2025 authorities issued warnings regarding the increase of sophisticated counterfeit lower-value wines entering mainstream supermarkets \cite{Telegraph_wine}. %(\url{https://www.bbc.com/news/articles/cvg3jzzjg3po}). 

Currently, wine producers primarily rely on external security measures on the bottle to mitigate counterfeiting, these include watermarks, holograms, microtexts, and barcodes on the labels, NFC (near-field communication) and RFID (radio frequency identification) tags, caps with tamper-proof mechanisms, and blockchain technologies \cite{maritano2024anti, soon2019developing, popovic2021novel, singh2021counterfeited, grolleau2022fine}. However, these controls focus solely on the packaging, not the wine itself, and have often proven insufficient \cite{lin2021fraud}.  Forgery and other fraudulent activities are dynamic and have become sophisticated enough to bypass these safeguards, heightening the ever-pressing need for rapid, cost-effective, and accurate methods for routine and on-site wine authentication \cite{sun2022real, lecat2017fraud}. The ability to authenticate wine without opening the bottle and sacrificing its contents would be a groundbreaking advancement, especially for high-end and small-batch wines, and is the challenge we address in this paper.

A myriad of analytical methods have been developed for wine authentication \cite{sun2022real,koljanvcic2024wine, mac2023current, popirdua2021review, pozo2024wine, arslan2021recent, rios2021spectralprint, ranaweera2021review, dos2017review }. Broadly, these methods can be categorised into three groups: 1)~classical instrumental analysis techniques including mass spectrometry and chromatography, 2) spectroscopic techniques and 3) other techniques such as DNA testing \cite{catalano2016experimental} and colourimetric sensing \cite{delima2020digital}. Classical instrumental analytical techniques such as high-performance liquid chromatography (HPLC), inductively coupled plasma mass spectrometry (ICP-MS), or isotope ratio mass spectrometry (IRMS) remain the most established and widely used approach for wine authentication \cite{koljanvcic2024wine, sun2022real}. While these methods offer high sensitivity and accuracy for wine analysis, they are inherently destructive. In contrast, spectroscopic techniques, often combined with multivariate statistical analysis, have proliferated recently due to their simplicity, speed, affordability and non-destructive nature \cite{popirdua2021review}. Consequently, non-invasive wine authentication logically points toward adopting non-destructive optical techniques. Spectroscopic techniques used for geographical and varietal authentication include nuclear magnetic resonance (NMR), UV/Vis, near- and mid-infrared (IR), Raman, and fluorescence spectroscopy \cite{pinto2025spectroscopic,ranaweera2021review, rios2021spectralprint, dos2017review}.

Fluorescence spectroscopy has proven to be a particularly powerful technique for wine analysis because of the natural presence of various fluorophores in wine. These fluorophores are mainly polyphenols such as phenolic acid, stilbene-like compounds and flavonoids, which provide valuable information about the terroir of wine \cite{rodriguez2001separation,airado2011front}. Fluorescence excitation-emission matrices (EEMs) have been used to quantify these fluorophores in wine \cite{cabrera2017front, dos2022direct,tarnok2023nonchromatographic} or to act as fingerprints to distinguish between wines based on grape variety \cite{azcarate2015modeling, dufour2006investigation, sadecka2020varietal, yin2009preliminary} or provenance \cite{airado2009usefulness, ranaweera2021authentication, ranaweera2021spectrofluorometric, suciu2019application, wang2024machine, wu2024geographical}. In all previous studies, however, obtaining the EEM remained an invasive method, requiring the wine bottle to be opened for analysis, and often included sample preparation steps such as filtration and dilution.

The primary challenge for through-bottle wine analysis arises from the green glass bottles commonly used for storing red wine, which produce complex spectroscopic signals of their own that obscure the signal from the wine itself. This has consequently limited research in this area, with only Harris \textit{et al.} using non-invasive NIR spectroscopy combined with machine learning models to assess the quality and origin of red wine \cite{harris2022non}. 

Spectroscopic analysis through a container has primarily been studied using a method called spatially offset Raman spectroscopy (SORS) \cite{mosca2021spatially}. In this technique, the excitation beam and the collection region are physically separated, allowing the collection of signals from deeper within the sample or from behind the container wall \cite{arroyo2021deep,lee2023direct}. SORS has diverse applications, such as identifying pharmaceuticals and drugs through packaging \cite{olds2011spatially}, non-invasive cancer diagnosis,  monitoring bone diseases \cite{nicolson2021spatially}, and analysing the compositions of paintings and other artefacts to assist art conservation \cite{conti2020advances}, to name a few. For wine analysis, SORS has been used to monitor the fermentation process of white wine through clear glass containers \cite{schorn2023spatially}. An alternative and more powerful Raman spectroscopy technique, related to SORS \cite{shillito2022focus}, uses an axicon lens to shape the excitation beam, enabling the acquisition of the Raman signal of whisky through the bottle while avoiding the signal of the clear glass \cite{fleming2020through}. 

In this paper, we demonstrate that an axicon-based geometry can be used to obtain the fluorescence spectra of red wine through its unopened coloured glass bottle, which would normally pose a significant confounding barrier to acquiring the signal from the wine itself. The axicon-based geometry shapes the excitation beam so that an annular beam is created on the surface of the bottle before focusing to a spot within the wine. The fluorescence emission from the wine is then collected through the centre `dark' part of the annular beam to avoid the fluorescence signal from the bottle. We also show, for the first time, that fluorescence spectra obtained with a single excitation wavelength of 785~nm are sufficient to successfully discriminate between twenty sealed wine bottles representing a wide range of red wine varietals. This non-invasive method, which does not require any sample preparation or scanning across multiple excitation wavelengths, holds promise for a fast, facile, and compact approach for wine authentication.

\section{Results and Discussion}
\subsection{Fluorescence system concept and characterisation}

\begin{figure*}[t!]
    \centering
    \includegraphics[width = \linewidth]{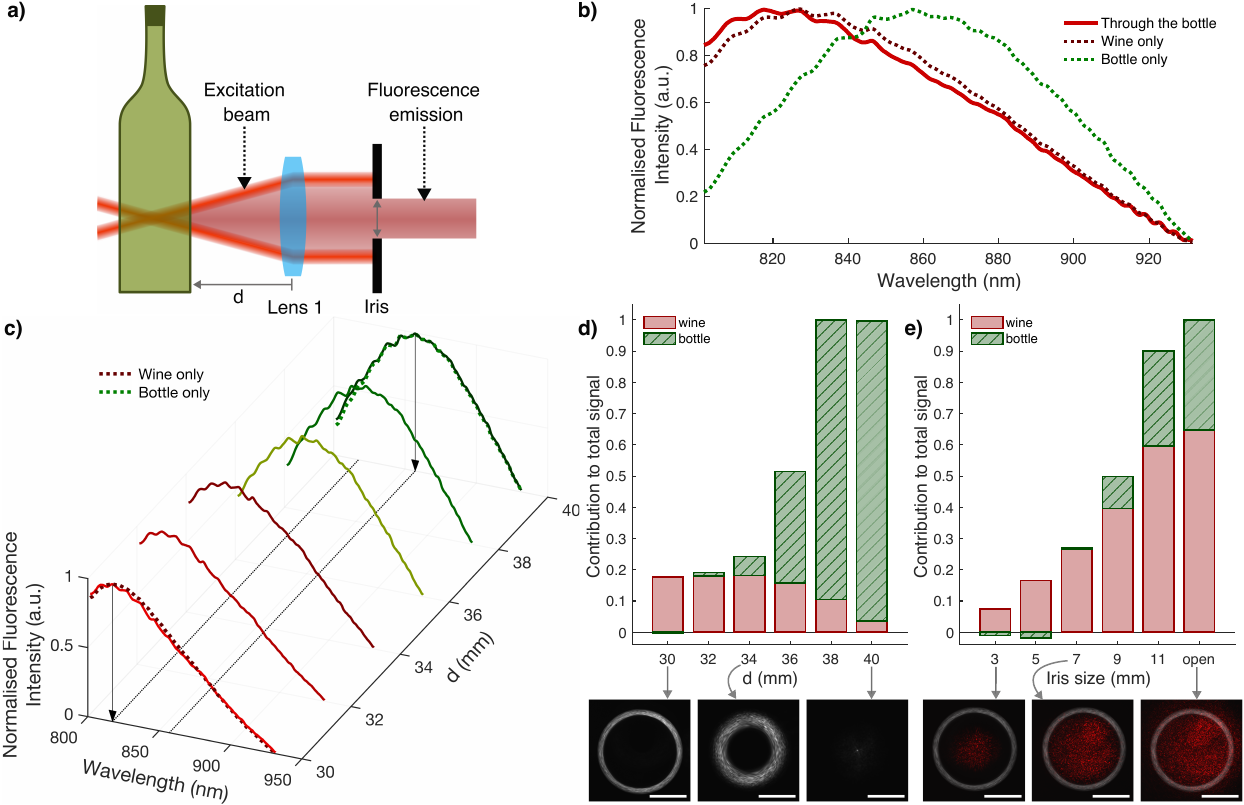}
    \caption{\textbf{Concept and characteristics of the through-bottle fluorescence system.} a) The fluorescence excitation beam (785~nm) is shaped so that an annular beam forms on the surface of the bottle, after which it focuses to a point inside the bottle. The fluorescence emission from the wine, excited at this focal point, is then collected through the centre of the annular excitation beam to avoid the fluorescence signal from the glass bottle. The iris blocks any fluorescence signal from the glass that is excited by the incident annular beam. Lens~1 has a focal length of 40~mm. b) The fluorescence spectrum of red wine (Shiraz~1) acquired through the bottle; the fluorescence spectra of only the bottle and only the wine are also plotted for reference. c) The fluorescence spectra measured as the wine bottle (Shiraz~2) is moved along the beam path. The spectra were normalised for clarity, and the fluorescence spectra of only the bottle and only the wine are plotted as dashed lines for reference. The fluorescence signal from the bottle is suppressed as the focus moves into the bottle. The relative contributions of the wine and the bottle to the total fluorescence signal are presented in the bar graphs for d) different bottle positions, and e) different iris diameters. For this wine bottle, the optimal system parameters are 30 mm from Lens~1 and an iris size of 7~mm. The insets show the excitation beam profile incident on the surface of the bottle (i.e. after being focused by Lens~1) in grey and the collection region in red. Scale bars are 1~mm.}
    \label{Fig: setup characteristics}
\end{figure*}

Our novel fluorescence technique relies on structuring the excitation beam into a cone-like shape, as depicted in Figure \ref{Fig: setup characteristics}a. This shape is obtained by exploiting the propagation properties of a beam created by an axicon lens, which results in an annular beam on the surface of the bottle and a bright focal point inside the sample. The fluorescence signal of the wine, excited at this focus, is then collected through the centre dark region of the annular beam, evading the fluorescence signal from the bottle. An iris is added to further block any stray fluorescence signal of the bottle, which is excited by the annular beam incident on the glass. See Materials and Methods for a detailed description and schematic of the complete optical setup used in this study. 

Figure \ref{Fig: setup characteristics}b shows the fluorescence spectrum of red wine measured through its bottle (``Through the bottle"), along with the fluorescence spectra of the empty glass bottle (``Bottle only") and the wine as measured in a quartz vial (``Wine only"). From these spectra, it is clear that the fluorescence spectrum obtained with our through-bottle system corresponds to the actual wine signal, while the fluorescence signal from the bottle is avoided.

To optimise the system and test its ability to obtain the fluorescence signal of red wine through the dark green glass bottles, two parameters were considered: 1) the distance $d$ between the bottle and Lens~1, i.e. how deep the beam focuses inside the bottle and 2) the iris aperture size. 

Figure \ref{Fig: setup characteristics}c shows the fluorescence spectra obtained as the distance $d$ between the sample (Shiraz~2) and Lens~1 (focal length of 40~mm) is varied. When the focus is on the surface of the bottle, at 40~mm, the signal is dominated by the fluorescence of the glass, peaking at 857~nm. As the bottle is moved closer to Lens~1, the beam focus shifts into the wine and the measured signal primarily reflects fluorescence from the wine itself, with a maximum at 817~nm. For this experiment, the iris diameter was kept constant at 7~mm. All spectra were normalised for clarity, and the fluorescence spectra of the wine alone (measured in a quartz vial) and the bottle alone were added for reference as dotted lines at 30~mm and 40~mm, respectively. 

The fluorescence intensity obtained with the through-bottle system ($I_{total}$) is mainly a combination of the signal from the wine and the glass bottle. This can be expressed as $I_{total} = a \times I_{wine} + b \times I_{bottle}$, where $I$ represents the fluorescence intensity, and $a$ and $b$ are the normalised weighting of the wine and bottle signal relative to the total signal. Since the wine bottle was opened for this experiment, the individual spectra $I_{wine}$ and $I_{bottle}$ were available, allowing $a$ and $b$ to be calculated for each through-bottle measurement by fitting a linear regression model. Figure \ref{Fig: setup characteristics}d shows the variation of these contributions with distance $d$. The insets show the profile of the excitation beam incident on the surface of the wine bottle at different positions. As the focus shifts into the wine, a drop in total fluorescence intensity is observed due to attenuation of the excitation light by the green glass. The signal from the bottle contributes to the total measured signal until the bottle is positioned 30~mm from Lens~1. At this point, the fluorescence signal of the wine is captured while avoiding the signal from the glass bottle. 

The iris diameter affects the size of the collection region of the fluorescence signal. Figure \ref{Fig: setup characteristics}e shows the effect of the iris size on the measured fluorescence intensity, and the relative contribution of the wine and the bottle to the total signal. The insets show images of the excitation beam profile on the surface of the bottle in grey (kept constant at 30~mm from Lens~1) and the fluorescence collection regions in red. The collection region was imaged by coupling an alignment laser into the collection fibre. The fluorescence signal from the bottle is avoided when the collection region does not overlap with the excitation beam incident on the bottle, i.e., when the collection cone is within the excitation cone. From Figure \ref{Fig: setup characteristics}e, an iris diameter of is 7~mm provides optimal performance, maximising the wine signal while suppressing the bottle contribution.

Red wine bottles are inconsistent with respect to colour, shape, and glass thickness. The photo in Figure~S1 of the Supplementary Information illustrates how much the glass properties differ from bottle to bottle. The glass thickness of six bottles was measured, ranging from 2.3~mm to 6.3~mm, with an average of 4.0~$\pm$~1.2~mm ($\pm$~$\sigma$). The optimal configuration of the system, therefore, changes slightly from sample to sample, depending on the fluorescence intensity of the wine, and the colour and thickness of the glass. 

\begin{figure*}[h!]
    \centering
    \includegraphics[width = \linewidth]{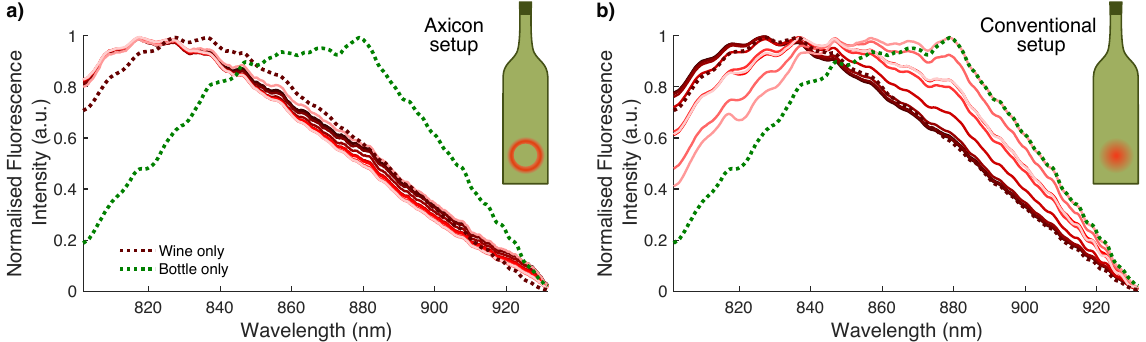}
    \caption{\textbf{Axicon and conventional Gaussian beam setup comparison.} The fluorescence spectra of red wine (Grenache~1) were measured through a bottle with a) the axicon setup and b) a conventional Gaussian beam setup. Ten spectra (solid lines) were taken at different positions around a single bottle. The spectra obtained with the axicon system were consistent and dominated by the fluorescence signal from the wine, whereas the spectra taken with a Gaussian beam varied depending on the bottle thickness. The fluorescence spectra solely of the wine and solely of the bottle are added for reference (dashed lines).}
    \label{Fig: Axicon vs Gaussian}
\end{figure*}

Conventionally, a standard Gaussian beam is used for fluorescence excitation. For comparison, the fluorescence spectrum of wine (Grenache~1) was measured through the bottle using both the cone-shaped excitation beam of the axicon setup and a Gaussian excitation beam, each in a back-scattering configuration. A schematic of the Gaussian beam setup is given in Figure~S2 in the Supplementary Information. With each setup, ten spectra were measured at different positions around a bottle, plotted in Figure \ref{Fig: Axicon vs Gaussian}. To illustrate the advantage of the axicon system, a bottle varying quite significantly in glass thickness (from 2.07~mm to 4.17~mm) was used for this experiment. With the axicon setup, a consistent fluorescence spectrum was obtained around the bottle, which is dominated by the signal from the wine. When a Gaussian beam was focused into the bottle, the measured fluorescence spectra varied quite significantly around the bottle; the fluorescence spectrum was sometimes dominated by the wine signal and sometimes by the signal from the glass bottle (see Figure \ref{Fig: Axicon vs Gaussian}b).

\subsection{Choice of excitation wavelength}

For the through-bottle fluorescence system, it is crucial that the excitation beam transmits through both the glass bottle and the wine. The excitation beam should not be absorbed by the bottle to ensure it reaches the sample, and it must also propagate into the wine deep enough to form a focus inside the liquid for fluorescence excitation. There is, however, a mismatch between the transmission of red wine and that of the wine bottles. 

\begin{figure*}[b!]
    \centering
    \includegraphics[width = \linewidth]{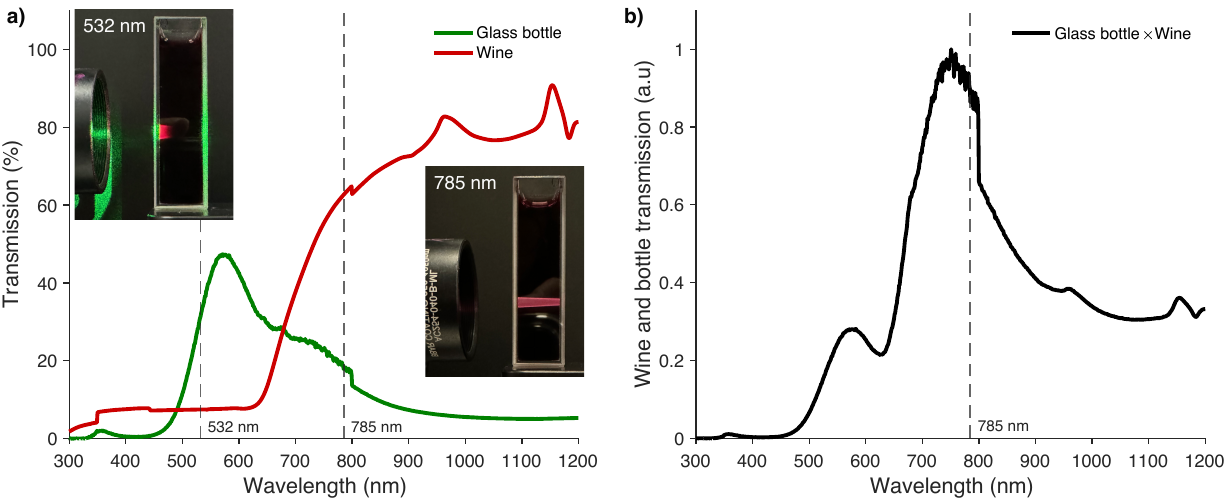}
    \caption{\textbf{Transmission metric to determine the optimal excitation wavelength for the through-bottle fluorescence system.} a) Transmission spectra of a typical red wine bottle and red wine (Shiraz). The insets show the propagation of a 532~nm and a 785~nm annular beam through red wine (Shiraz~2); 100~mW laser power used for excitation. For the through-bottle system to work, the excitation beam must be transmitted through the green glass and the wine to create a focal point inside the wine for fluorescence excitation. b) A plot of the transmission spectrum of the glass bottle~$\times$ the transmission spectrum of the wine. The peak position can be used to identify the optimal excitation wavelength for the through-bottle setup; the chosen excitation wavelength of 785~nm is indicated on the plot.}
    \label{Fig: transmission metric}
\end{figure*}

The transmission spectra of red wine (Shiraz) and a typical Shiraz wine bottle are presented in Figure~\ref{Fig: transmission metric}a, showing that when the transmission through the bottle is high, the transmission of the wine is low, and vice versa. The insets show the propagation of an annular beam through red wine (Shiraz~2) created with a 532~nm and a 785~nm laser source. The propagation of the excitation beam through the sample could be visualised by the fluorescence emission from the wine. Since the wine exhibits high absorption at 532~nm, the annular beam does not propagate deep into the sample, preventing it from forming a focus inside the wine. In contrast, due to the high transmission of 785~nm light through wine, the excitation beam at this wavelength propagates through the sample, allowing for the formation of a focal point inside the wine necessary for through-bottle measurements.

An optimal excitation wavelength for the through-bottle system can be identified by determining where the product of the transmission spectra of the glass bottle and that of the wine is at a maximum, see Figure~\ref{Fig: transmission metric}b. A peak around 750~nm is observed for this metric. The chosen wavelength of 785~nm falls within the peak and is also widely used for Raman excitation, meaning dedicated laser sources and spectrometers for 785~nm are readily available. Depending on the bottle colour and glass thickness, the transmission of the 785~nm excitation beam through wine bottles can range from 10\% to 40\%, with an average of 22\%~$\pm$~10\% ($\pm$~$\sigma$). Red wine exhibits weaker fluorescence emission with near-IR excitation sources compared to near-UV/Vis sources that are commonly used in fluorescence studies of wine \cite{ranaweera2021authentication}. As a result, detecting the fluorescence at 785~nm excitation requires a sensitive spectrometer, such as the one used in this study. 

To assess the potential photobleaching of the wine due to the excitation laser, the fluorescence emission of 4~mL of red wine (in a clear vial) was measured under continuous illumination of a focused 785~nm laser at a power of 420~mW for an exposure time of 20~min. None of the three wines exhibited a significant decrease in fluorescence intensity or change in their emission profile under these harsh conditions (see Figure S3 in the Supplementary Information). Given that the through-bottle system operates on a larger volume of wine (750~mL), with a shorter exposure time (approximately 25~s, corresponding to 5 scans each with a 5~s integration time), and a lower laser power of (approximately 70~mW reaching the wine after transmission through the bottle), no photodamage to the wine is expected from the excitation laser.

\subsection{Wine discrimination and classification}

\begin{figure*}[b!]
    \centering
    \includegraphics[width = \linewidth]{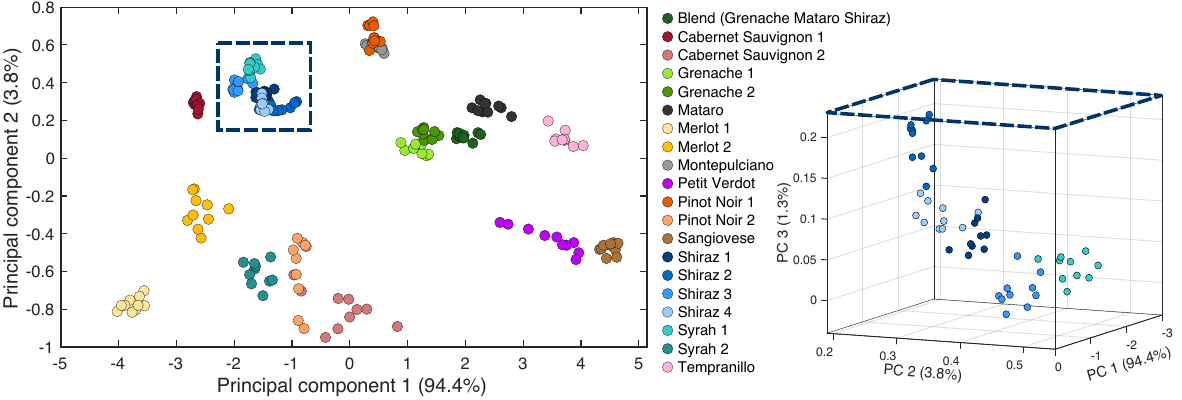}
    \caption{\textbf{Principal component (PC) plot discriminating between twenty different unopened wines.} This PC plot shows the grouping of 10 spectra of each wine, with the different varietals tending to group together. The first three principal components describe 99.4\% of the variation. The inset on the right shows that the five Shiraz/Syrah varietals that clustered together can be further separated when considering the third PC. }
    \label{Fig: PCA}
\end{figure*}

The fluorescence spectra of twenty unopened wines of varying varietals were measured through their native bottle. The spectra were analysed using principal component analysis (PCA). PCA is a well-known multivariate technique used to map the data onto a new set of orthogonal axes, known as principal components (PCs) that explain most of the variance in the underlying data \cite{abdi2010principal}. The PC plot, shown in Figure \ref{Fig: PCA}, therefore helps visualise groups and trends within a data set. The first two PCs describe 98.1\% of the variance in the sample. The ten fluorescence spectra taken of each wine bottle show significant grouping, confirming the repeatability of the measurements around a wine bottle. Excellent separation between different wines in the PC plot proves that a single fluorescence spectrum (at 785~nm excitation) is sufficient to discriminate between the different wines. 

Linear discriminant analysis (LDA) is a multivariate linear classification algorithm. In contrast to PCA, which is unsupervised, LDA is a supervised method that seeks to maximise the separation between predefined classes. LDA was applied to classify the twenty bottles of wine based on their fluorescence spectra, where training was done on 40\% of the spectra ($n_{train} = 80$) and the remaining 60\% used for testing ($n_{test} = 120$). 100\% of the wines were correctly classified (confusion matrix shown in Figure S4 in the Supplementary Information). 

The ability to use a single excitation source to obtain a fingerprint of red wine is a surprising result, as nearly all fluorescence studies on wine typically use a range of excitation wavelengths (usually between 245~nm and 500~nm) to generate EEMs for analysis \cite{airado2009usefulness, airado2011front, cabrera2017front, coelho2015fluorescence, ranaweera2021spectrofluorometric, rios2024usefulness, dos2022direct}. An exception is the study by Dufour \textit{et al.}, which used one fluorescence excitation (250 – 350~nm) and one emission (275 – 450~nm) spectrum to examine the variety and typicality of red wines \cite{dufour2006investigation}. 

%To the best of our knowledge, this study of Dufour \textit{et al.} in 2006 is also the first to propose using fluorescence for obtaining fingerprints of wine. 

As previously mentioned, Raman spectroscopy commonly uses 785~nm as an excitation source. In Raman applications, a fluorescence background is unwanted, as it can mask the weaker Raman signals. Consequently, due to the high fluorescence of red wine, Raman spectroscopy of wine is mostly conducted on white wines, which exhibit much lower fluorescence \cite{magdas2018wine, magdas2019testing, martin2015raman}. For Raman studies on red wine, a 1064~nm excitation source has been used to reduce the fluorescence \cite{mandrile2016controlling} or when using 785~nm, considerable care is taken to reduce the fluorescence background either through sample preparation techniques \cite{fuller2021alcoholic, qu2020chemical} or computationally \cite{gallego2010rapid}. A study using Raman spectroscopy (785~nm excitation) to measure phenolic compounds in white and red wine, found that they could distinguish between different varietals of the red wine \cite{deneva2019using}. They note, however, that the difference among varietals is attributed to the fluorescence background rather than the Raman peaks. Here, we confirm that the fluorescence spectrum of red wine, excited at 785~nm, can serve as a fingerprint of the wine, as well as carry information about the wine varietal. 

In Figure \ref{Fig: PCA}, wines of the same varietals tend to group together as can be seen for Shiraz/Syrah (blue markers), and Grenache (green markers) varietals. The individual wines are, however, still sufficiently separable if the third PC is considered as shown in the inset. It is worth noting that Shiraz and Syrah wines are the same grape varietal, only differing in winemaking styles. Additionally, the Grenache (50\%), Mataro (30\%), and Shiraz (20\%) blend is situated between the single varietal Grenache and Mataro wines on the PC plot. There is, therefore, a correlation between the PCs of the spectra and the varietal of the wine, showing potential to use this method for varietal identification. Notably, three wines do not group with their other varietals as expected: Cabernet Sauvignon 2, Pinot Noir 2 and Syrah 2. It was observed that in the case of Cabernet Sauvignon 2, its bottle was much darker/thicker than the others; for the other two wines, the fluorescence signal from the wine itself was low. For these three wines, the bottle signal contributed more to the total signal and could explain why they grouped together. When the fluorescence spectra of these wines were measured through lighter bottles, they grouped much closer with their other varietals (PC plot shown in Figure~S5 in the Supplementary Information) confirming that the wine/bottle ratio influenced these measurements. This issue, i.e. very dark bottles or low fluorescing wines, can be overcome by using an excitation laser with a higher output power.

\begin{figure}[htb]
    \centering
    \includegraphics[width = 0.5\linewidth]{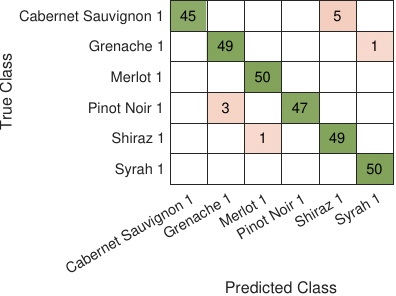}
    \caption{\textbf{Classification of the fluorescence spectra of 6 wines acquired through different bottles.} The LDA model was trained on the spectra measured through one of the bottles ($n_{train} = 60$). The confusion matrix shows the model performance on a test set consisting of the spectra measured through the five remaining bottles ($n_{test} = 300$). Only ten spectra were misclassified, giving a success rate of 96.7\%.}
    \label{Fig: LDA}
\end{figure}

To test how much the bottle influences the measured fluorescence spectrum of the wine, the fluorescence emission was measured for six different wines decanted into six different bottles (6 wines $\times$ 6 bottles $\times$ 10 measurements around each bottle $= 360$ spectra in total). An LDA model was trained on the spectra measured in the first bottle ($n_{train} = 60$) and the spectra measured through the five remaining bottles were used for testing ($n_{test} = 300$). Figure \ref{Fig: LDA} displays the confusion matrix of the LDA results, indicating that 96.7\% of the spectra were correctly classified. This demonstrates that the through-bottle fluorescence system can accurately identify a wine even when it is not in its own bottle. While LDA offers a simple and interpretable approach for classification and performs well on our dataset, more advanced machine learning methods, such as neural networks, could be explored in future work, particularly with larger datasets \cite{wang2024machine, lu2023identification, lee2024learning, armstrong2023machine}.

\section{Conclusion}

The fluorescence spectrum of red wine was successfully measured through the bottle by making use of an axicon lens to shape the excitation beam. This approach allows for the detection of auto-fluorescence from the unopened wine while eliminating the signal from the green glass bottle. We identified an optimal configuration for the measurements, considering the focus position inside the bottle and the iris aperture size, to maximise the fluorescence signal but still suppress the bottle signal. The axicon system outperformed a conventional setup that uses a standard Gaussian beam for excitation. A fluorescence excitation wavelength of 785~nm was selected since it can be transmitted through both the bottle and the wine; we also demonstrated that this excitation laser does not cause photodamage to the wine. The PCA results of the through-bottle fluorescence spectra, excited at 785~nm, showed excellent separation of twenty unopened wine bottles, with the potential to use these spectra for varietal classification. Using LDA, a 100\% accurate classification was achieved for the twenty wines. We continue to be able to identify wines when they are decanted into different wine bottles, with an accuracy of 96.7\%. These results lay the groundwork for a compact, simple, and rapid device capable of non-invasive wine authentication in unopened bottles, with potential applications extending to other food and drug safety fields.

\section{Materials and Methods}
\subsection{Samples}

Twenty different Australian red wines were included in this study. The wine bottles were either provided by the Australian Wine Research Institute or purchased from national retail outlets. The details of each wine such as the varietal, vintage, and region are reported in Table S1 in the supplementary information.

\subsection{The optical setup}

\begin{figure*}[b!]
    \centering
    \includegraphics[width = 0.75\linewidth]{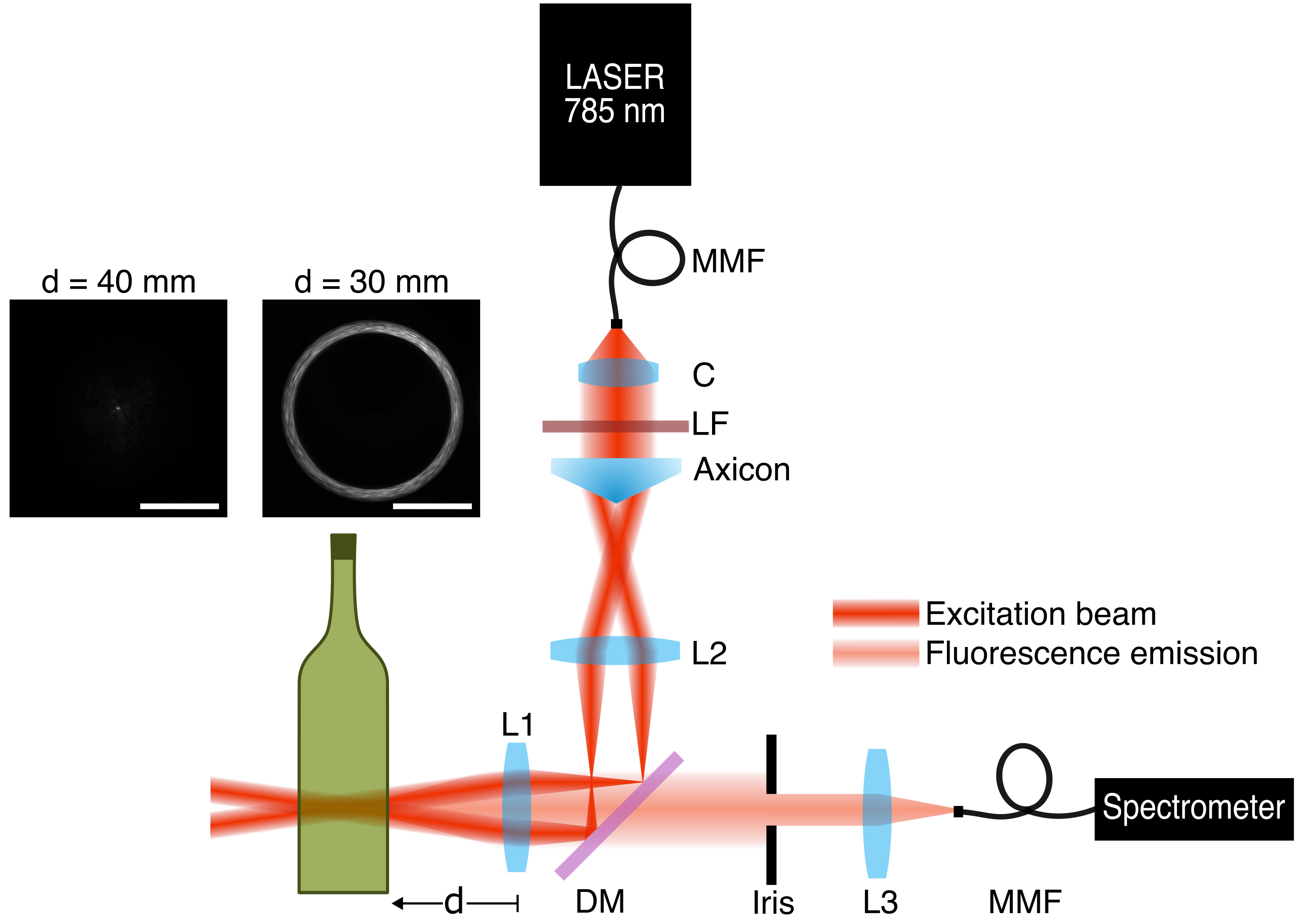}
    \caption{\textbf{A schematic of the system used to measure the fluorescence spectra of red wine through the bottle.} The insets show the beam profiles created by the axicon lens at two distances from L1: at 30~mm an annular beam is created which collapses to a focused spot at 40~mm. MMF: multi-mode fibre; C: collimator; LF: line filter for 785 nm; DM: dichroic mirror; L1: lens (40~mm); L2: lens (100~mm) and L3: lens (50~mm). Scale bars are 1~mm.}
    \label{Fig: Through-bottle fluo setup}
\end{figure*}

The optical setup used in this study is schematically shown in Figure \ref{Fig: Through-bottle fluo setup}. The setup is similar to that used by Fleming \textit{et al.} for Raman spectroscopy measurements of alcoholic spirits through transparent bottles \cite{fleming2020through}. A 785 nm laser (WP785, Wasatch Photonics) was used for fluorescence excitation. The collimated beam, after being filtered by a line filter (LL01-785, Semrock), passes through an axicon lens (AX255-B, $5^{\circ}$  angle, Thorlabs) to create a Bessel beam. A 4-f system (lenses L1 and L2) is used to deliver the Bessel excitation beam to the sample plane. A dichroic mirror (LPD02- 785RU, Semrock) is used to separate the excitation beam from the fluorescence signal. An adjustable iris is used to suppress the fluorescent signal from the bottle before the remaining collected light is delivered to the spectrometer (WP785, Wasatch Photonics).

Lens L2 performs a Fourier transform, converting the Bessel beam into an annular beam at its focal plane, and the second lens L1 reverts the annular beam into a Bessel beam at the focal plane of lens L1 (40 mm), creating the cone-shaped excitation beam. The chosen focal lengths of the lenses are based on the optimal conditions determined by Shillito \textit{et al.} for Raman spectroscopy through 70~mm diameter clear glass whisky bottles using an axicon geometry  \cite{shillito2022focus}; red wine bottles typically have a diameter of 75~mm. Images of the beam created at 30~mm and 40~mm after L1 are shown as insets in Figure \ref{Fig: Through-bottle fluo setup}: the beam still takes on an annular shape at 30~mm, after which the beam collapses to a focused spot at 40~mm. To spatially separate the fluorescent signal from the wine and the fluorescence signal from the bottle, the wine bottle is placed so that the annular beam is incident on the surface of the bottle and the focus is formed inside the wine bottle. The fluorescence signal from the bottle which is excited by the annular beam incident on the glass is eliminated by incorporating an iris before the spectrometer.

A multi-mode fibre, rather than a single-mode fibre, was used to deliver the laser beam to the setup to maximise the power of the excitation beam. However, this means that a perfect Bessel beam is not created at the focus and a speckle pattern is seen on the annular beams. The beam is also distorted by the glass of the bottle and the wine itself, however, a focus with sufficient intensity is created to excite the auto-fluorescence of the wine.  

The laser power of the Bessel beam is 360~mW before entering the bottle, however only about 22\% ($\sim$~70~mW) of this laser power is transmitted through the green glass of the bottle to the wine sample. The amount of laser power delivered to the sample differs from bottle to bottle since red wine bottles can significantly vary in colour, shape, and glass thickness.

This system design provides a simple, compact measurement geometry where the excitation and collection paths are collinear. By comparison, a SORS geometry requires physical separation of the excitation and collection paths \cite{mosca2021spatially, shillito2022focus}.

\subsection{Fluorescence measurements}
All fluorescence measurements were taken with an iris diameter of 7~mm unless otherwise indicated. The optimal distance of the bottle from L1 was slightly different for each bottle (between 20–30~mm). The height of the excitation beam was approximately 5~cm from the bottom of the wine bottle. Ten measurements at different positions around the bottle were taken for each wine; the label and glass seams of the bottle were avoided. Each fluorescence spectrum was recorded using the Wasatch Photonics ENLIGHTEN Spectroscopy Software. Each acquired spectrum was an average of 5 spectra with an integration time of 5~s. The integration time was reduced for wines of higher intensity to avoid saturating the detector; the power was kept constant at 360~mW at the sample plane. 

\subsection{Spectral processing}
All processing and analysis were performed using MATLAB (R2023b). The measured spectra were normalised between 0 and 1 after a dark subtraction and were subsequently smoothed. Standard multivariate analysis algorithms of PCA and LDA were employed for the classification of the wines based on the spectral features of the fluorescence emission. 

\subsection{Transmission measurements}
All transmission spectra were recorded using an Agilent Cary 5000 UV-Vis-NIR Spectrophotometer. For analysis of the wine bottles, 40 mm × 20 mm sections of glass were cut from the bottles, and the transmission was measured through these sections. A 1~cm quartz cuvette was used for all red wine samples.

\section{Acknowledgements}
The authors acknowledge early contributions to the experimental setup from Erik Schartner and would like to thank Alexander Trowbridge, Kwang Jun Lee, and Dion Turner for valuable discussions. This work was performed in part at the University of Adelaide OptoFab hub of the Australian National Fabrication Facility, utilising Commonwealth and SA State Government funding, for preparing the glass sections. The authors acknowledge funding support from an ARC Laureate Fellowship (grant FL210100099), AWRI-UA Collaborative Research Investment Fund (University of Adelaide), and the University of Adelaide Research Scholarship (119858).

\section{Author contributions}
KD conceived and supervised the study. AK built the experimental setup, performed the measurements, and analysed the data with input from RPM. EW provided specialist expertise regarding wine characterisation. AK drafted the paper with contributions from RPM, GDB, and KD. All authors critically reviewed and approved the manuscript.

\FloatBarrier

\renewcommand{\thefigure}{S\arabic{figure}}
\renewcommand{\theequation}{S\arabic{equation}}
\renewcommand{\thetable}{S\arabic{table}}
\setcounter{figure}{0}
\setcounter{table}{0}
\setcounter{equation}{0}

\section*{\huge\centering Supplementary Information}
% \subsection*{Details of wine samples}

% The varietal, vintage, wine name, region and winery of the twenty bottles of wine used in this study are reported in Table \ref{table: wine samples}. 

\begin{table}[htb]
\caption{Details of the wine samples used in this study.}
\begin{adjustbox}{width=1\textwidth}
\label{table: wine samples}
\begin{tabular}{llllllll}
\multicolumn{1}{l}{} & \multicolumn{1}{l}{\textbf{Sample Name}} & \multicolumn{1}{l}{\textbf{Wine Name}} & \multicolumn{1}{l}{\textbf{Company/Winery}} & \multicolumn{1}{l}{\textbf{Varietal}} & \multicolumn{1}{l}{\textbf{Vintage}} & \multicolumn{1}{l}{\textbf{Region}} & \multicolumn{1}{l}{\textbf{State}} \\ \hline
1 & Blend (GMS) & Avatar & Teusner & \begin{tabular}[c]{@{}l@{}}Grenache (50\%), Mataro\\(30\%), Shiraz (20\%)\end{tabular} & 2021 & Barossa Valley & SA \\ \hline
2 & Cabernet Sauvignon 1 & Cabernet Sauvignon & Moss Wood & 88\% Cabernet Sauvignon & 2017 & Margaret River & WA \\ \hline
3 & Cabernet Sauvignon 2 & Reserve & Chateau Yering & Cabernet Sauvignon & 2014 & Yarra Valley & VIC \\ \hline
4 & Grenache 1 & Warboys Vineyard & Angove & Grenache & 2019 & McLaren Vale & SA \\ \hline
5 & Grenache 2 & The Nook & Buller Wines & Grenache & 2021 & Victoria & VIC \\ \hline
6 & Mataro & Resurrection & Langmeil & Mataro & 2020 & Barossa & SA \\ \hline
7 & Merlot 1 & Merlot & 3drops & Merlot & 2018 & Great Southern & WA \\ \hline
8 & Merlot 2 & Merlot & Tempus Two & Merlot & 2022 & \begin{tabular}[c]{@{}l@{}}Multi-Regional New \\ South Wales Blend\end{tabular} & NSW \\ \hline
9 & Montepulciano & Montepulciano & The Ethereal One & Montepulciano & 2021 & Adelaide Hills & SA \\ \hline
10 & Petit Verdot & Petit Verdot & Pirramimma & Petit Verdot & 2020 & McLaren Vale & SA \\ \hline
11 & Pinot Noir 1 & Inverness Ridge & Yering Station & Pinot Noir & 2017 & Yarra Valley & VIC \\ \hline
12 & Pinot Noir 2 & Pinot Noir & Clandestine Vineyards & Pinot Noir & 2021 & Adelaide Hills & SA \\ \hline
13 & Sangiovese & \begin{tabular}[c]{@{}l@{}}The New Australian \\ Collection\end{tabular} & Coriole & Sangiovese & 2022 & McLaren Vale & SA \\ \hline
14 & Shiraz 1 & Fordwich Hill & Margan & Shiraz & 2018 & Hunter Valley & NSW \\ \hline
15 & Shiraz 2 & Estate Label & Taylors & Shiraz & 2021 & \begin{tabular}[c]{@{}l@{}}Limestone Coast /\\ Clare Valley\end{tabular} & SA \\ \hline
16 & Shiraz 3 & The Hatchling & Schubert & Shiraz & 2020 & Barossa Valley & SA \\ \hline
17 & Shiraz 4 & Stablemate & Sidewood & Shiraz & 2021 & Adelaide Hills & SA \\ \hline
18 & Syrah 1 & Macclesfield & Longview & Syrah & 2019 & Adelaide Hills & SA \\ \hline
19 & Syrah 2 & \begin{tabular}[c]{@{}l@{}}Jardwadjali Land \\ Syrah\end{tabular}& \begin{tabular}[c]{@{}l@{}}Little Brunswick \\ wine co.\end{tabular} & Syrah & 2021 & Grampians & VIC \\ \hline
20 & Tempranillo & Tempranillo & Hither and Yon & Tempranillo & 2022 & McLaren Vale & SA 
\end{tabular}
\end{adjustbox}
\end{table}

% \subsection*{Transmission of red wine}

% The transmission spectra of wine bottles were measured. The analysis focused on dark green glass bottles commonly used for red wines. In order to analyse the glass of red wine bottles, 40 mm $\times$ 20 mm sections of glass were cut from five different wine bottles. Figure \ref{Fig: Glass transmission} shows the transmission spectra of the glass cutouts. A maximum transmission and the wavelength used in this study is highlighted on this graph at around 568~nm and 785~nm, respectively. The transmission spectra were obtained with a Agilent Cary 5000 UV-Vis-NIR Spectrophotometer.

% The transmission of light through the glass depends not only on the wavelength and the specific bottle but also on the thickness and curvature of the glass. The variability of the transmission over one of these sections of glass (Bottle 4) was tested by measuring the power of a 785 nm laser beam before and after the glass at different points and angles. Transmission varying between 14\% - 19\% were observed for the section of glass. The transmission varies quite significantly over the small section of glass, therefore, even more variation is expected over a single wine bottle considering the thickness and curvature difference of the neck, shoulder and body of a typical wine bottle.

% \subsection*{The conventional Gaussian beam setup}

% The through-bottle Gaussian beam setup used in this study is schematically shown in Figure \ref{Fig: Gauss setup}. 
\FloatBarrier
\begin{figure}[h]
    \centering
    \includegraphics[width = 0.7\linewidth]{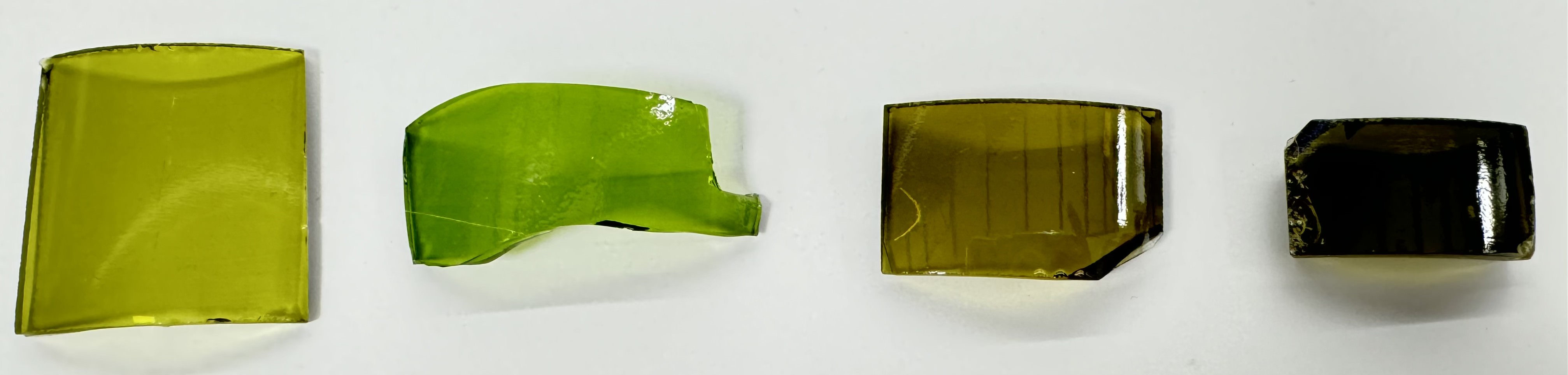}
    \caption{A photo of four glass pieces from wine bottles demonstrating the variation in colour typically observed in red wine bottles. }
    \label{Fig: Glass pieces}
\end{figure}

\begin{figure}
    \centering
    \includegraphics[width = 0.6\linewidth]{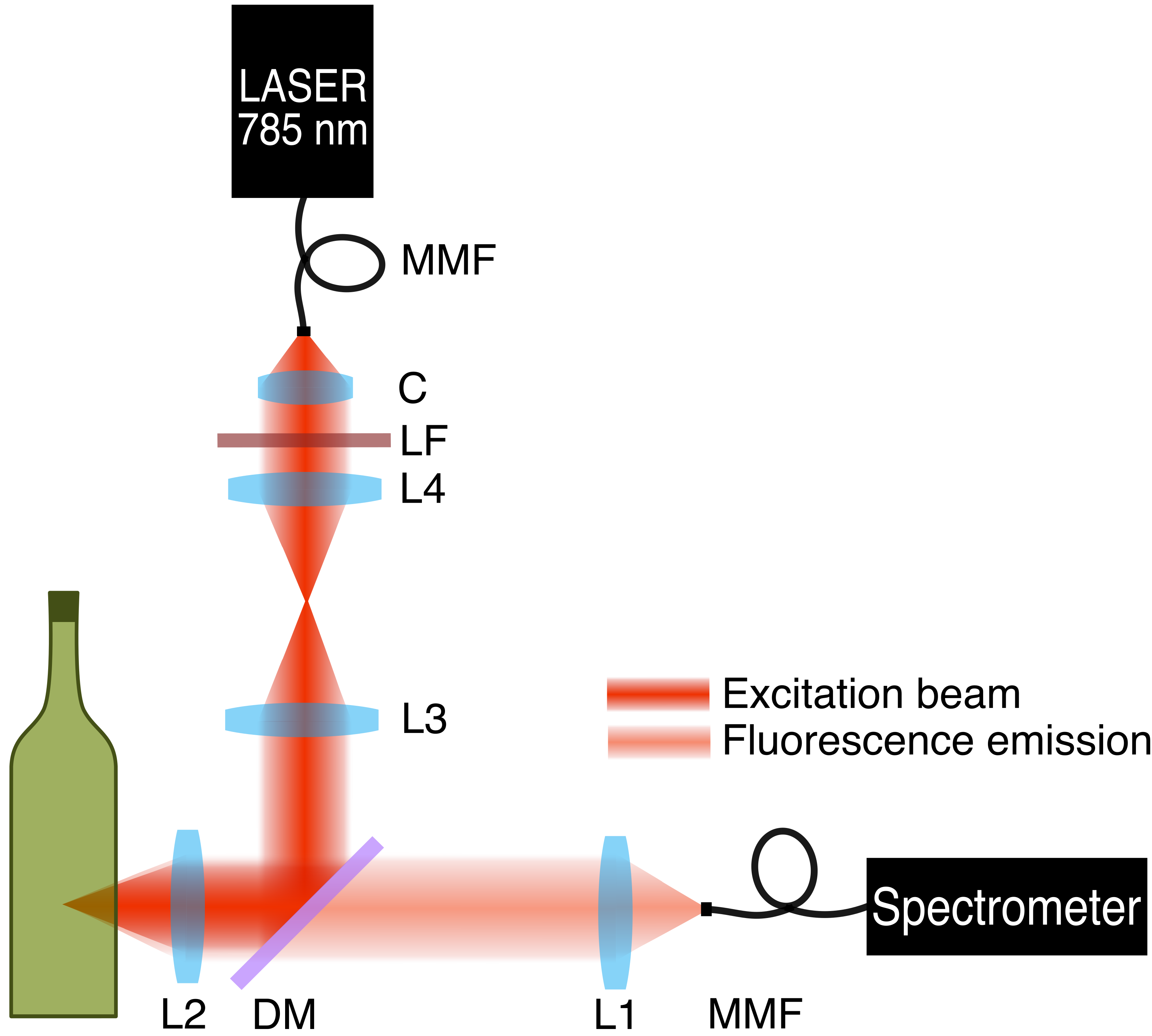}
    \caption{A schematic of the system used to measure the fluorescence spectra of red wine through the bottle with a Gaussian excitation beam. This setup is identical to what is shown in the main text, but with the axicon replaced by a plano-convex lens (L4) with a focal length of 60~mm which then delivers a Gaussian beam to the sample. A Gaussian beam is therefore incident on the surface of the bottle after which it focuses inside the bottle. An iris is also not needed in this setup, since the excitation beam and collection region overlap. MMF: multi-mode fibre; C: collimator; LF: line filter for 785 nm; DM: dichroic mirror; L1: lens (50~mm); L2: lens (40~mm) and L3: lens (100~mm); L4: lens (35~mm).}
    \label{Fig: Gauss setup}
\end{figure}

% \begin{figure*}
%     \centering
%     \includegraphics[width = 0.7\linewidth]{Figures/TransmissionImages.png}
%     \caption{The propagation of an annular beam through red wine (Shiraz 2) created with a) a 532~nm and b) a 785~nm laser source. The propagation of the excitation beam through the sample is visualised by the fluorescence emission of the wine. Since the wine exhibits high absorption of light at 532~nm, the annular beam does not propagate far into the sample and no focus forms inside the sample. Due to high transmission of 785~nm light through wine, the excitation beam at this wavelengths travels through the sample allowing for the formation of a focal point in the wine necessary for through-bottle measurements. Power of excitation beams: 100~mW.}
%     \label{Fig: Wine transmission}
% \end{figure*}

% \FloatBarrier
% \subsection*{Wine classification}
% The through-bottle setup was used to measure the fluorescence spectra of 20 different unopened wine bottles. These spectra were classified using linear discriminant analysis (LDA) by splitting the spectra into a training (40\% of spectra, $n_{train} = 80$) and a test set (60\% of spectra, $n_{test} = 120$). The confusion matrix of the LDA results are presented in Figure \ref{Fig: All 20 bottles} , showing 100\% correct classification of the 20 wines.

% \subsection*{Photostability}

\begin{figure}
    \centering
    \includegraphics[width =1\linewidth]{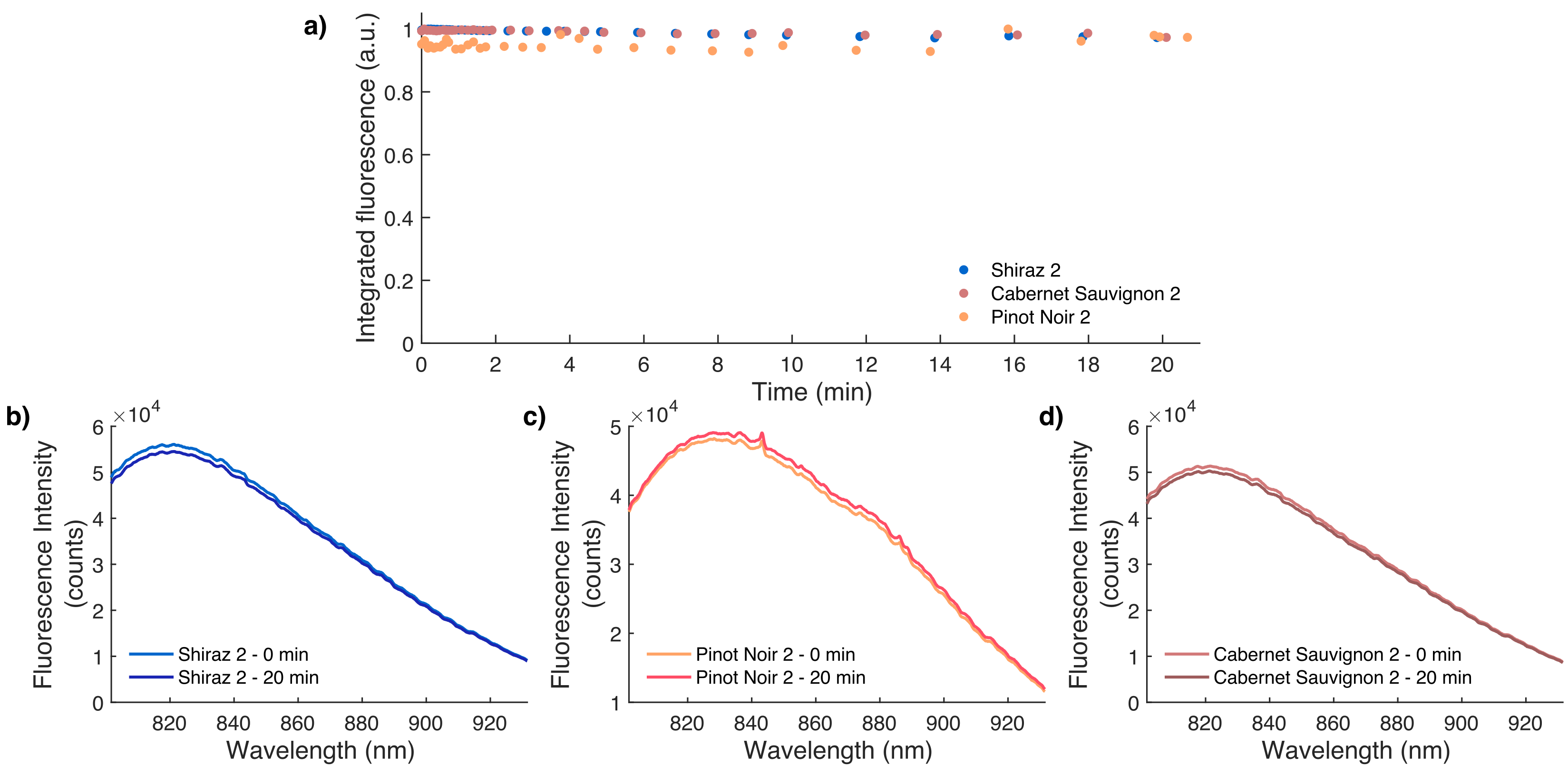}
    \caption{The fluorescence spectra of three wines were measured (in a 4~mL quartz cuvette) under continuous illumination for 20 min with a 785~nm laser at 420 mW. a) The integrated fluorescence intensity of each wine was tracked over time, with intensities normalised to their initial values at 0 min for clarity. No photobleaching was observed in any of the samples. b)~–~d) show the fluorescence spectra of the three wines at both the start (0 min) and end (20 min) of exposure, confirming no change in fluorescence profile even after prolonged laser illumination.}
    \label{Fig: Stability}
\end{figure}

\begin{figure*}
    \centering
    \includegraphics[width = 0.9\linewidth]{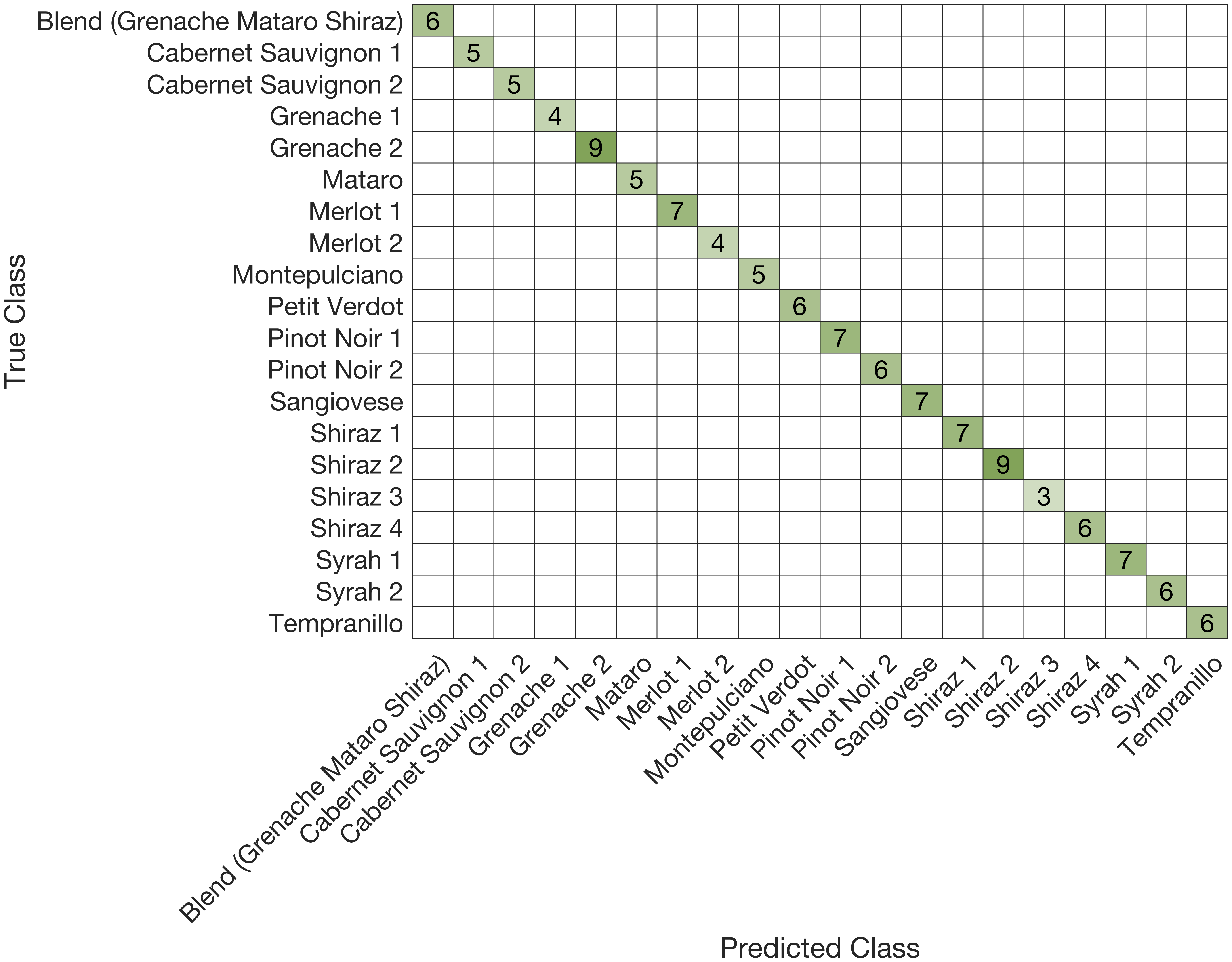}
    \caption{The through-bottle setup was used to measure the fluorescence spectra of 20 different unopened wine bottles. These spectra were classified using linear discriminant analysis (LDA) by splitting the spectra into a training (40\% of spectra, $n_{train} = 80$) and a test set (60\% of spectra, $n_{test} = 120$). The confusion matrix of the LDA results presented here shows the 100\% correct classification of the 20 wines.}
    \label{Fig: All 20 bottles}
\end{figure*}

% \FloatBarrier

% \subsection*{Wine discrimination}
% The principal component (PC) plot of the fluorescence spectra of 20 unopened wine bottles measured with the through-bottle setup, shows that wines of the same varietals tend to group together except for Cabernet Sauvignon 2, Pinot Noir 2 and Syrah 2. It was observed that in the case of Cabernet Sauvignon 2, its bottle was much darker/thicker than the others; for the other two wines, the fluorescence signal from the wine itself was low. For these three wines, the bottle signal contributed more to the total signal and could explain why they grouped together. The fluorescence spectra of these wines were measured through different bottles: a lighter/thinner wine bottle (Cabernet Sauvignon 1) and clear white wine bottles (Pinot Noir 2 and Syrah 2). These spectra grouped much closer with their other varietals as shown in Figure \ref{Fig: PCA with outliers} confirming that the wine/bottle ratio influenced these measurements.

\begin{figure*}
    \centering
    \includegraphics[width = \linewidth]{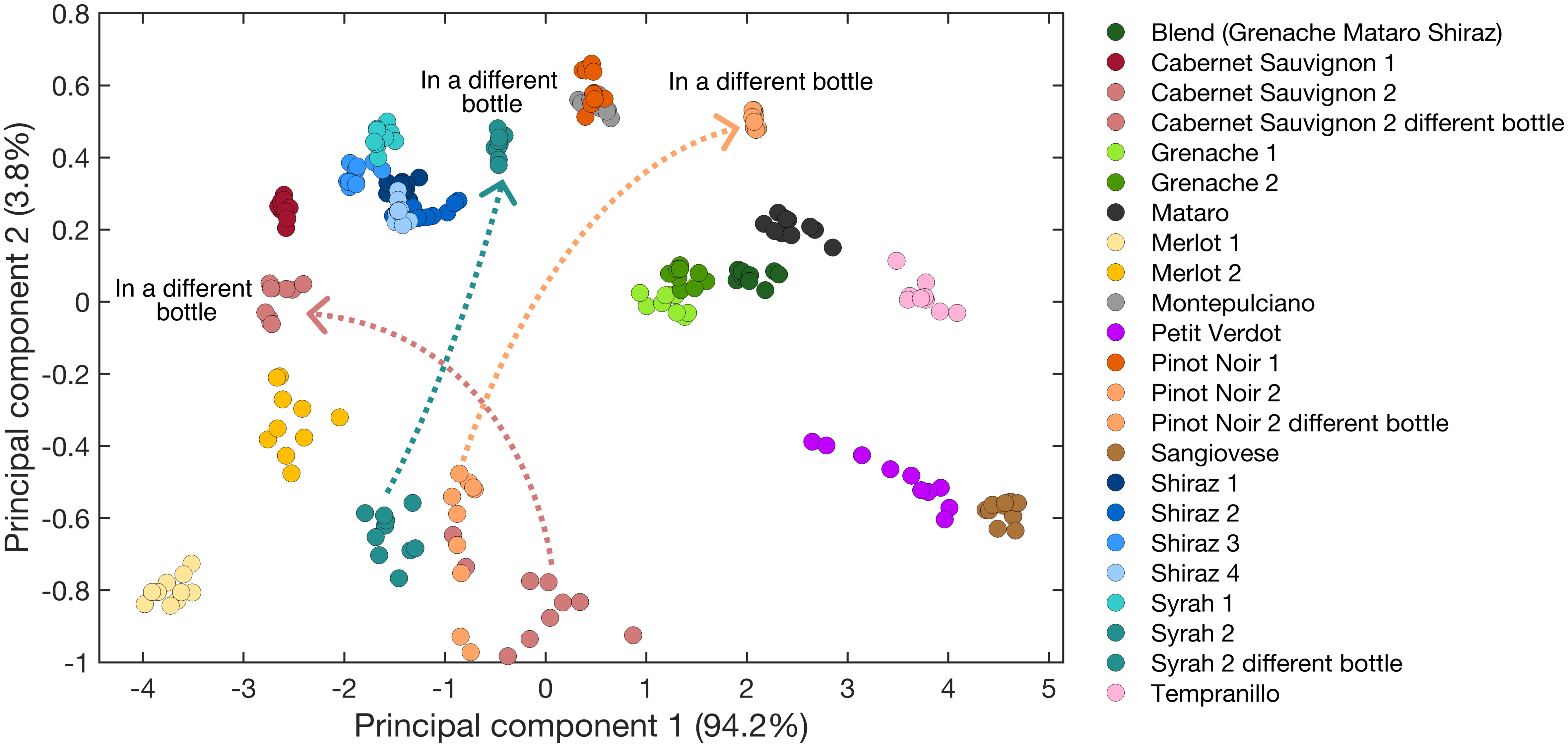}
    \caption{The principal component (PC) plot of the fluorescence spectra of 20 unopened wine bottles measured with the through-bottle setup. This plot shows that wines of the same varietals tend to group together, except for Cabernet Sauvignon 2, Pinot Noir 2 and Syrah 2. It was observed that in the case of the former, its bottle was much darker/thicker than the others; and for the latter two wines, the fluorescence signal from the wine itself was low. For these three wines, the bottle signal contributed more to the total signal and could explain why they grouped together. These three wines did not initially group with their varietal as expected, however, once the spectra were measured through a different lighter bottle (for Cabernet Sauvignon 1) or through a clear bottle (for Pinot Noir 2 and Syrah 2), they grouped much closer to their corresponding varietal (follow the arrows), confirming that the wine/bottle ratio influenced these measurements.}
    \label{Fig: PCA with outliers}
\end{figure*}

% \FloatBarrier
% \bibliographystyle{unsrt}

% \bibliography{ReferencesAK.bib}

% \end{document}

\end{document}